\documentclass[prx,aps,superscriptaddress,nofootinbib,twocolumn]{revtex4-1}

\usepackage{mathrsfs}
\usepackage{amsfonts}
\usepackage{amssymb}
\usepackage{amsmath}
\usepackage{graphicx}
\usepackage[usenames,dvipsnames]{color}
\usepackage[colorlinks=true,citecolor=blue,linkcolor=magenta]{hyperref}
\usepackage{ulem}
\usepackage{lmodern}

\newcommand{\figpath}{.}

\newcommand{\Tr}{\mathrm{Tr}}

\newcommand{\abs}[1]{| #1 |}

\newcommand{\ket}[1]{\vert{ #1 }\rangle}

\begin{document}

\title{Fault-tolerant fermionic quantum computation based on color code}

\author{Ying Li}

\affiliation{Graduate School of China Academy of Engineering Physics, Beijing 100193, China}
\affiliation{Department of Materials, University of Oxford, Parks Road, Oxford OX1 3PH, United Kingdom}

\date{\today}

\begin{abstract}
An important approach to the fault-tolerant quantum computation is protecting the logical information using the quantum error correction. Usually, the logical information is in the form of logical qubits, which are encoded in physical qubits using quantum error correction codes. Compared with the qubit quantum computation, the fermionic quantum computation has advantages in quantum simulations of fermionic systems, e.g.~molecules. In this paper, we show that the fermionic quantum computation can be universal and fault-tolerant if we encode logical Majorana fermions in physical Majorana fermions. We take a color code as an example to demonstrate the universal set of fault-tolerant operations on logical Majorana fermions, and we numerically find that the fault-tolerance threshold is about $0.8\%$.
\end{abstract}

\maketitle

\section{Introduction}

Quantum computation can solve some problems much faster than classical computation~\cite{Nielsen2010}, e.g.~simulating large quantum systems. Because quantum states are fragile, fault tolerance is crucial to quantum computation. There are two approaches to the fault-tolerant quantum computation: error-correction-based quantum computation~\cite{Shor1996} and topological quantum computation~\cite{Nayak2008}. In the error-correction-based quantum computation, the quantum information is encoded using quantum error correction codes and protected by actively detecting and correcting errors. In the topological quantum computation, the quantum information is encoded in the state of non-Abelian anyons and processed by braiding anyons, and these braiding operations can tolerate small perturbations.

Majorana fermions are non-Abelian anyons considered as candidates for the topological quantum computation~\cite{Nayak2008}, which have been observed in recent experiments~\cite{Mourik2012, Mebrahtu2013, NadjPerge2014, Xu2015, Haim2015, Albrecht2016, Deng2016, He2017}. Majorana fermions can also be used for implementing the fermionic quantum computation, in which fermionic modes instead of qubits are the basic units that carry the quantum information~\cite{Bravyi2002}. The fermionic quantum computation is polynomially equivalent to the qubit quantum computation but has advantages in quantum simulations of fermionic systems, e.g.~molecules~\cite{Wecker2014, Poulin2015, Reiher2016, Kassal2011}. By encoding each qubit in four Majorana fermion modes~\cite{Bravyi2006}, the fermionic quantum computation can efficiently simulate the qubit quantum computation. However, to simulate fermions using qubits, fermions need to be encoded in the state of the entire qubit system as a whole, and some local operations on fermions are realised by qubit operations on a significant portion of total qubits~\cite{Bravyi2002, Ortiz2001, Whitfield2011, Jones2012, Hutter2015, Seeley2012, Tranter2015}. These non-local qubit operations cost resources and may slow down the quantum computing.

Quantum computation based on Majorana fermions still needs the quantum error correction~\cite{Bravyi2010, Terhal2012, Vijay2015, Landau2016, mypaper, Plugge2016, Vijay2017, Litinski2017}. Braiding Majorana fermions does not provide all the operations required by the universal quantum computation, therefore some topologically unprotected operations have to be introduced to complete the operation set~\cite{Bravyi2006}. These unprotected operations could cause errors, which need to be corrected using the quantum error correction. Majorana fermions may also suffer change-tunnelling errors caused by the fermionic bath in the environment, e.g.~unpaired electrons in the superconductor~\cite{Goldstein2011, Trauzettel2012, Cheng2012, Loss2012, Campbell2015}. In order to correct errors, we can either encode physical qubits in Majorana fermions~\cite{Bravyi2006} and then encode logical qubits in physical qubits~\cite{mypaper, Plugge2016, Litinski2017} or directly encode logical qubits in Majorana fermions~\cite{Bravyi2010, Terhal2012, Vijay2015, Landau2016, Vijay2017}. In either case, the logical machine is a qubit quantum computer. In this paper, we propose using logical Majorana fermions encoded in physical Majorana fermions to realise the genuine fermionic quantum computation. We demonstrate that the universal operation set can be implemented on logical Majorana fermions using local operations in a fault-tolerant manner.

Color code is a class of topological quantum error correction codes~\cite{Bombin2006}. We can construct codes for encoding logical Majorana fermions using color codes~\cite{Bravyi2010}. Taking a triangular color code as an example, we demonstrate the universal set of operations on logical Majorana fermions, and we also propose an efficient decoding algorithm for the code based on the code unfolding~\cite{Bombin2012, Kubica2015}. The error-rate threshold is an important measure of the code performance: only when the error rate is lower than the threshold, the fault-tolerant quantum computation can be realised. Usually, color codes provide much lower thresholds~\cite{Landahl2011} than high-threshold codes, e.g.~the surface code. By numerically simulating the quantum error correction, we find that the threshold of the Majorana fermion color code is about $0.8\%$ for entangling operations, which is comparable to the threshold of the surface code~\cite{Wang2011}.

The paper is organised as follows. Codes for encoding logical Majorana fermions in physical Majorana fermions are introduced in Sec.~\ref{sec:code}. We introduce the color code in Sec.~\ref{sec:color}. The universal fault-tolerant fermionic quantum computation is discussed in Sec.~\ref{sec:universal}, Sec.~\ref{sec:operations} and Sec.~\ref{sec:array}. The decoding algorithm is described in Sec.~\ref{sec:correction}, and the numerical result about the fault-tolerance threshold is shown in Sec.~\ref{sec:threshold}. A summary is given in Sec.~\ref{sec:summary}.

\section{Majorana fermion codes for encoding logical Majorana fermions}
\label{sec:code}

Majorana fermions are described by Hermitian operators $c_i$ obeying $\{c_i, c_j\} = 2\delta_{i,j}$, and each operator corresponds to a mode of Majorana fermions.

Majorana fermion codes are stabiliser codes of Majorana fermions~\cite{Bravyi2010}. Many quantum error correction codes, e.g.~Calderbank-Shor-Steane codes, color codes~\cite{Bombin2006}, and the surface code~\cite{Dennis2002} are stabiliser codes of qubits~\cite{Nielsen2010}. For a code of $2n$ physical Majorana fermion modes, the dimension of the Hilbert space is $2^n$. The logical information is encoded in a subspace defined by the stabiliser group $\langle S_p \rangle$, in which each generator is a product of Majorana fermion operators $S_p = i^{\frac{1}{2}\abs{V_p}} \prod_{i\in V_p} c_i$, where $V_p$ is the set of modes supporting $S_p$. We remark that changing the order of operators in the product may change the overall sign. The number of Majorana fermion modes $\abs{V_p}$ in a stabiliser generator is always even, otherwise the generator changes the parity of the number of fermions, which is not physical. These stabiliser generators $\{S_p\}$ are Hermitian and mutually commutative. The phase $i^{\frac{1}{2}\abs{V_p}}$ makes sure that $S_p$ is Hermitian. The number of common modes $\abs{V_p \cap V_{p'}}$ for any two stabiliser generators is even, so that two generators $S_p$ and $S_{p'}$ are commutative. Because $S_p^2 = 1$, a stabiliser generator $S_p$ has two eigenvalues $+1$ and $-1$. The logical subspace is the subspace that all stabiliser generators take the eigenvalue $+1$, i.e.~$S_p\ket{\psi} = \ket{\psi}$ if the state $\ket{\psi}$ is in the logical subspace. If there are $k$ independent stabiliser generators, the dimension of the logical subspace is $2^{n-k}$, and there are total $2^k$ subspaces corresponding to $2^k$ sets of eigenvalues. In general, we can choose any of these subspaces as the logical subspace. Errors may bring the state out of the logical subspace, and they are detected by checking whether the state is still in the logical subspace and, if the state is not in the logical subspace, in which subspace the state ends up, i.e.~measuring eigenvalues of stabiliser generators.

The logical state is described using logical operators. Logical operators are also products of Majorana fermion operators, and they commute with all stabiliser generators. For a logical operator $L = \eta \prod_{i\in V_L} c_i$ defined on the set of modes $V_L$, $\abs{V_L \cap V_p}$ is always even for any stabiliser generator $S_p$, so that $[L, S_p] = 0$. Here, $\eta$ is a phase. For example, we can encode one qubit in four Majorana fermion modes~\cite{Bravyi2006} $c_1$, $c_2$, $c_3$ and $c_4$. The only stabiliser operator is $c_1c_2c_3c_4$, and logical operators are $\sigma^{\rm x} = ic_1c_4$ and $\sigma^{\rm z} = ic_3c_4$, which are Pauli operators of the encoded qubit.

We can encode logical Majorana fermions in qubits. However, such an approach does not lead to the genuine fermionic quantum computation. Majorana fermions can be encoded using the Jordan-Wigner transformation $c_{2i-1} = \sigma^{\rm x}_i \prod_{j<i} \sigma^{\rm z}_j$ and $c_{2i} = \sigma^{\rm y}_i \prod_{j<i} \sigma^{\rm z}_j$~\cite{Ortiz2001, Hutter2015}, where the product is taken over all qubits with a number less than $i$. Therefore, sometime we need to perform a sequence of qubit operations across the entire system to operate one Majorana fermion mode. For example, to perform the operation $U_{i,j} = \exp[\theta (\tilde{c}_{2i-1}\tilde{c}_{2j}-\tilde{c}_{2i}\tilde{c}_{2j-1})]$, we need to perform about $2\abs{i-j}$ gates on qubits between $i$ and $j$~\cite{Whitfield2011}. Although these gates can be implemented in a constant time by using ancillary qubits~\cite{Jones2012}, any other operation on fermionic modes from $i$ to $j$ cannot be performed in parallel with $U_{i,j}$, because corresponding qubits are in use. In the worst case, the entire computer is occupied to operate only two fermionic modes, e.g.~$U_{1,N}$, where $N$ is the total number of fermionic modes. We remark that the locality of the encoding of individual fermionic modes could be improved by using the Bravyi-Kitaev transformation~\cite{Bravyi2002, Seeley2012, Tranter2015}. Fermionic modes can also emerge in qubit topological codes with twists~\cite{Bombin2010, Brown2017}, therefore, are protected by the code.

We can encode one logical Majorana fermion in an independent set of physical Majorana fermions~\cite{Bravyi2010}, so that each logical Majorana fermion can be operated independently to realise the genuine fermionic quantum computation. We consider a code of $2n$ Majorana fermion modes with $n-1$ independent stabiliser generators, and the mode $c_{2n}$ is not included in any stabiliser operator, i.e.~$2n \notin \bigcup_p V_p$. We remark that stabiliser generators should cover all other $2n-1$ modes, so that any single-mode error can be detected. For such a code, $\bar{c}  = i\prod_{i=1}^{2n-1} c_i$ is a logical operator. Because $\abs{V_p}$ are all even, $\bar{c}$ commutes with all stabiliser operators $S_p$. This logical operator satisfies $\bar{c}^\dag = \bar{c}$ and $\bar{c}^2 = 1$. For any Majorana fermion operator $c$ other than these $2n-1$ operators $c_1,\ldots,c_{2n-1}$, we can find that the logical operator satisfies $\{ \bar{c}, c \} = 0$. If we consider two copies of the code, two logical operators $\bar{c}_1$ and $\bar{c}_2$ also satisfy $\{ \bar{c}_1, \bar{c}_2 \} = 0$. Therefore, the logical operator $\bar{c}$ represents a logical Majorana fermion mode.

The mode $c_{2n}$ is not included in neither stabiliser generators nor the logical operator, so it is redundant and can be removed, i.e.~we can encode one logical Majorana fermion mode in $2n-1$ physical Majorana fermion modes. We cannot define the Hilbert space of $2n-1$ Majorana fermion modes, therefore we cannot define the logical subspace of the code directly. After introducing the $2n$-th mode, the Hilbert space of the total $2n$ modes is $2^n$-dimensional, and the logical subspace is $2$-dimensional. It does not matter whether the $2n$-th mode is a physical Majorana fermion or a logical Majorana fermion. Therefore, for even number of logical Majorana fermion modes, we can define their common logical subspace.

\section{Color code}
\label{sec:color}

\begin{figure}[tbp]
\centering
\includegraphics[width=1\linewidth]{\figpath 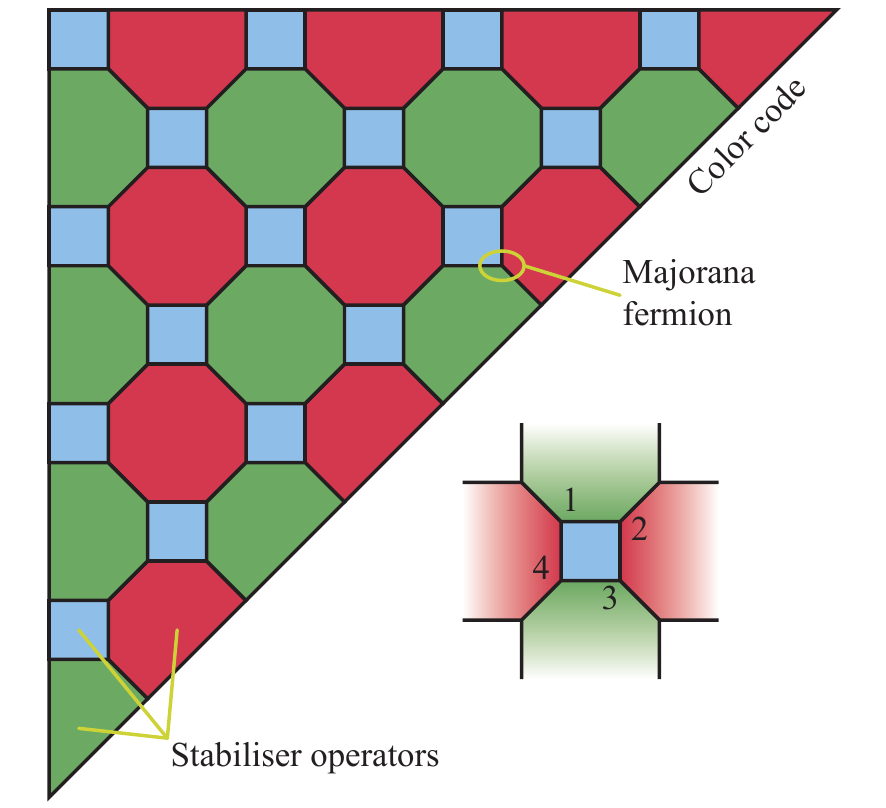}
\caption{
Triangular code based on the $(4,8^2)$ lattice. Each vertex represents a Majorana fermion mode, and each plaquette represents a stabiliser generator. The side length (the number of vertices on the side) of the lattice is $d = 9$.
}
\label{fig:code}
\end{figure}

Color code is a class of topological stabiliser codes~\cite{Bombin2006}. A color code is defined on a lattice, e.g.~the lattice in Fig.~\ref{fig:code}, in which plaquettes can be colored with three colors (i.e.~{\it green}, {\it blue} and {\it red}), and every two plaquettes sharing an edge have different colors. The number of vertices on any plaquette is always even, and the number of vertices shared by any two plaquettes is also always even. Therefore, we can construct stabiliser generators of Majorana fermions according to the color code lattice: a Majorana fermion mode is placed at each vertex, and the product of Majorana fermion operators on a plaquette is a stabiliser generator~\cite{Bravyi2010}. In other words, for each $S_p$, the label $p$ denotes a plaquette, and $V_p$ denotes the set of vertices (i.e. Majorana fermion modes) on the plaquette. Because of properties of the lattice, these stabiliser generators satisfy that $\abs{V_p}$ and $\abs{V_p \cap V_{p'}}$ are even. The code encodes one logical Majorana fermion mode if the total number of vertices is odd, e.g.~triangular codes obtained by removing a vertex from a color code lattice defined on a sphere~\cite{Bombin2006} such as the code in Fig.~\ref{fig:code}. The logical operator $\bar{c}$ is the product of all Majorana fermion operators.

In the following, we will take the color code in Fig.~\ref{fig:code} as an example to discuss how to perform fault-tolerant fermionic quantum computation. This code is based on the $(4,8^2)$ lattice but different from the $(4,8^2)$ triangular code reported in Ref.~\cite{Bombin2006}. We are interested in this lattice, because we find that such a lattice provides a high error-rate threshold in the fault-tolerant qubit quantum computation based on Majorana fermions~\cite{mypaper}.

\section{Universal fermionic quantum computation}
\label{sec:universal}

\begin{figure*}[tbp]
\centering
\includegraphics[width=1\linewidth]{\figpath 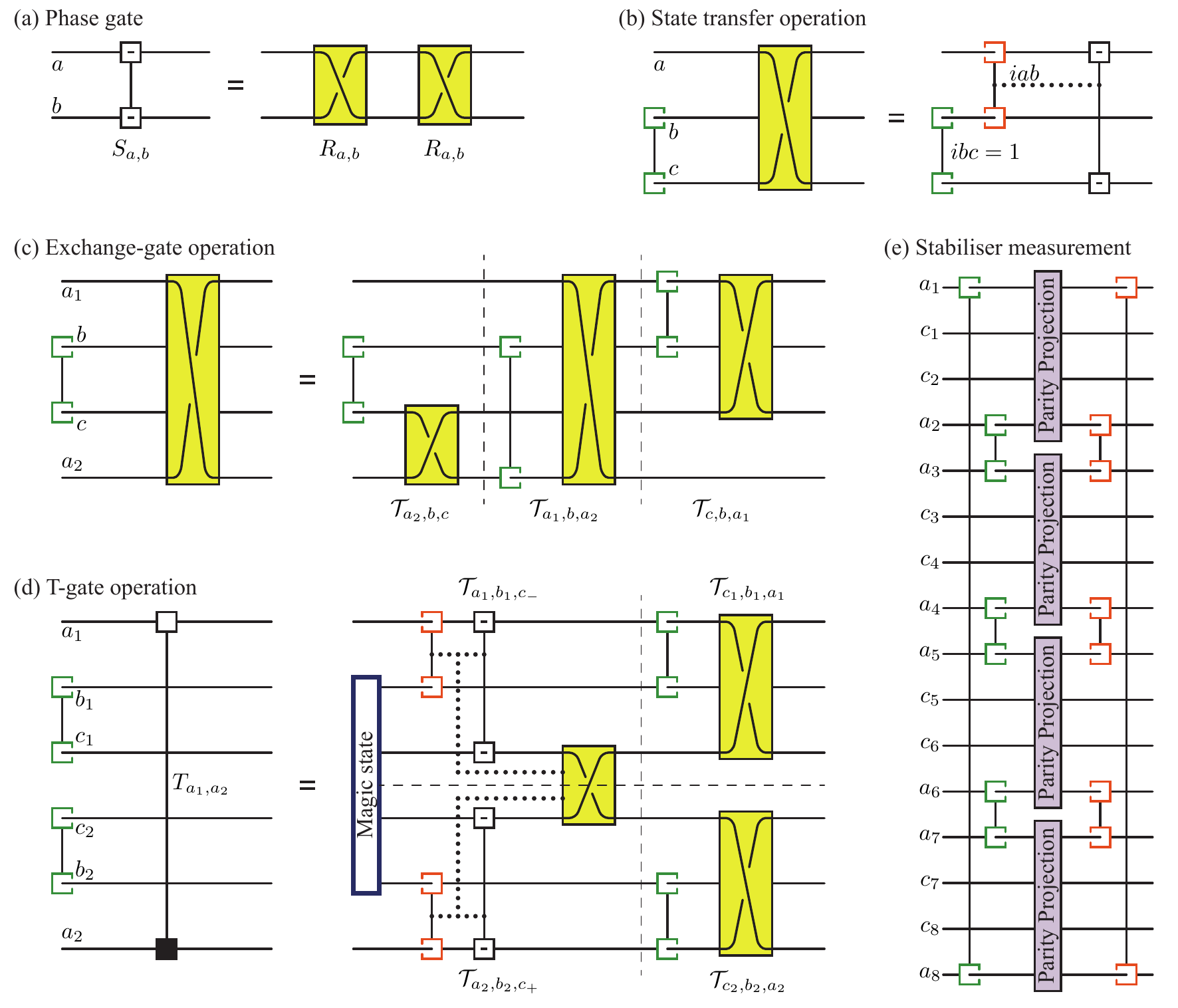}
\caption{
Circuits for Majorana fermions. Each horizontal line represents a Majorana fermion mode.
(a) A phase gate $S_{a,b}$ is equivalent to two exchange gates $R_{a,b}$.
(b) The state transfer operation $\mathcal{T}_{a,b,c}$ transfers the state of $a$ to $c$. Green boxes represent initialisation operations, and red boxes represent measurement operations. The phase gate $S_{a,c}$ depends on the measurement outcome of $iab$ (indicated by the dotted line).
(c) The exchange gate $R_{a_1,a_2}$ realised using a sequence of state transfer operations.
(d) The gate $T_{a_1,a_2}$ realised using the magic state and state transfer operations. The magic state is the eigenstate with $ib_1c_- = 1$ and $ib_2c_+ = 1$, where $c_\pm = \frac{1}{\sqrt{2}}(c_1 \pm c_2)$. Outcome-dependent phases gates in $\mathcal{T}_{a_1,b_1,c_-}$ and $\mathcal{T}_{a_2,b_2,c_+}$ are equivalent to the outcome-dependent exchange gate $R_{c_2,c_1}$ (indicated by dotted lines), which can be realised using the circuit in (c).
(e) The circuit for measuring the stabiliser operator of eight Majorana fermion modes $c_1, c_2, \ldots, c_8$ using ancillary Majorana fermion modes $a_1,a_2,\ldots,a_8$. Each pair of ancillary modes is initialised in the eigenstate $ia_{2i}a_{2i+1} = 1$ ($a_9 = a_1$). Parity projections are performed on each group of four Majorana fermion modes, and outcomes are $c_{2i-1}c_{2i}a_{2i-1}a_{2i} = \upsilon_{2i-1,2i}$. The measurement outcome of the stabiliser operator is $c_1c_2\cdots c_8 = -\upsilon_{1,2}\upsilon_{3,4}\cdots \upsilon_{7,8}$. After parity projections, ancillary modes are measured, and outcomes are $ia_{2i}a_{2i+1} = \eta_{2i,2i+1}$. To complete the stabiliser measurement, phase gates need to be performed according to these outcomes as shown in Table~\ref{table}.
}
\label{fig:circuits}
\end{figure*}

The operation set allowing universal fermionic quantum computation includes i) the initialisation of a pair of Majorana fermion modes in the eigenstate $ic_1c_2 = 1$; ii) the measurement of $ic_1c_2$; iii) the non-destructive measurement of $c_1c_2c_3c_4$, which is called {\it parity projection} in this paper; iv) the exchange gate $R_{c_1,c_2} = \frac{1}{\sqrt{2}}(1+c_1c_2)$, and v) the gate $T_{c_1,c_2} = \exp(\frac{\pi}{8} c_1c_2)$~\cite{Bravyi2002}. Any parity-preserving unitary operator can be composed using these operations.

In this section, we show that the operation set is still universal if we replace the exchange gate $R$ with the phase gate $S_{c_1,c_2} = c_1c_2$ and the $T$ gate with the preparation of the magic state. The magic state of the $T$ gate is an eigenstate of four Majorana fermion modes with $i\frac{1}{\sqrt{2}}(c_1+c_2)c_4 = 1$ and $i\frac{1}{\sqrt{2}}(c_1-c_2)c_3 = 1$. Considering the code for encoding one qubit in four Majorana fermion modes (i.e.~$c_1c_2c_3c_4 = 1$, $\sigma^{\rm x} = ic_1c_4$ and $\sigma^{\rm z} = ic_3c_4$), such an eigenstate can be written as $\ket{\rm A} = \frac{1}{\sqrt{2}}(\ket{0} + e^{i\frac{\pi}{4}}\ket{1})$, which is the magic state for the qubit non-Clifford gate $T =  \exp(i \frac{\pi}{8} \sigma^{\rm z})$~\cite{Bravyi2005}. In the next two sections, we demonstrate how to implement the universal operation set, including the initialisation, measurement, parity projection, phase gate and magic-state preparation, on logical Majorana fermions.

We can realise the exchange gate using the phase gate provided that the initialisation and measurement are available, and vice versa. The exchange gate exchanges states of two Majorana fermion modes and adds a phase to one of them, i.e.~$R_{a,b}aR_{a,b}^\dag = -b$ and $R_{a,b}bR_{a,b}^\dag = a$. By exchanging two Majorana fermion modes twice, the phase is added to each of them, which is equivalent to a phase gate $S_{a,b} = R_{a,b}^2$ ($S_{a,b}aS_{a,b}^\dag = -a$ and $S_{a,b}bS_{a,b}^\dag = -b$) [see Fig.~\ref{fig:circuits}(a)]. As we will show next, the state of a Majorana fermion mode can be transferred up to a random phase using the initialisation and measurement. Therefore, the exchange gate can be realised by firstly exchanging states of two Majorana fermion modes using the initialisation and measurement and then adjusting the phase using the phase gate.

The state transfer operation is described by the superoperator $\mathcal{T}_{a,b,c} = \mathcal{M}_{a,b,c} \mathcal{I}_{b,c}$ [see Fig.~\ref{fig:circuits}(b)]. Here, $\mathcal{I}_{b,c} = \sum_{\nu = \pm 1} [b^\frac{1-\nu}{2}] [\pi^{(ibc)}_\nu]$ is the initialisation that prepares $b$ and $c$ in the eigenstate $ibc = 1$, and $\mathcal{M}_{a,b,c} = \sum_{\mu = \pm 1} [(ac)^\frac{1+\mu}{2}] [\pi^{(iab)}_\mu]$ is the measurement of $iab$ followed by a phase gate on $a$ and $c$ (i.e.~$S_{a,c} = ac$) depending on the measurement outcome $iab = \mu$. We note that $\pi^{(ic_1c_2)}_\eta = \frac{1}{2}(1+\eta ic_1c_2)$ is a projector to the eigenstate $ic_1c_2 = \eta$, and $[U]\rho = U\rho U^\dag$ is a superoperator. Using $(ac)^\frac{1+\mu}{2} \pi^{(iab)}_\mu b^\frac{1-\nu}{2} \pi^{(ibc)}_\nu = \frac{1+ac}{2} b^\frac{1-\nu}{2} \pi^{(ibc)}_\nu $, we get $\mathcal{T}_{a,b,c} = [R_{a,c}] \mathcal{I}_{b,c}$. Therefore, the state transfer operation $\mathcal{T}$ is equivalent to an initialisation operation on $b$ and $c$ followed by an exchange gate on $a$ and $c$, i.e.~the state of $a$ is transferred to $c$ with an additional phase. The phase is determined because of the outcome-dependent phase gate. Here, we have assumed that the measurement is non-destructive. If the measurement is destructive, $b$ and $c$ may not be initialised in the correct eigenstate, but the sate of $a$ can still be transferred to $c$ with the correct phase.

The exchange gate $S_{a_1,a_2}$ can be realised using a sequence of state transfer operations $\mathcal{T}_{c,b,a_1}\mathcal{T}_{a_1,b,a_2}\mathcal{T}_{a_2,b,c}$ [see Fig.~\ref{fig:circuits}(c)]. Such a combination of state transfer operations is equivalent to $[R_{a_1,a_2}] \mathcal{I}_{b,c}$, i.e.~an initialisation operation on $b$ and $c$ followed by an exchange gate on $a_1$ and $a_2$. Here, we have used that $R_{c,a_1}R_{a_1,a_2}R_{a_2,c} = R_{a_1,a_2}$.

The gate $T_{a_1,a_2}$ can also be realised using a sequence of state transfer operations. The overall operation is $\mathcal{T}_{c_2,b_2,a_2}\mathcal{T}_{c_1,b_1,a_1}\mathcal{T}_{a_1,b_1,c_-}\mathcal{T}_{a_2,b_2,c_+}$ [see Fig.~\ref{fig:circuits}(c)], where $c_\pm = \frac{1}{\sqrt{2}}(c_1 \pm c_2)$. Initialisation operations $\mathcal{I}_{b_1,c_-}$ and $\mathcal{I}_{b_2,c_+}$ (in $\mathcal{T}_{a_1,b_1,c_-}$ and $\mathcal{T}_{a_2,b_2,c_+}$, respectively) prepare the magic state on $b_1$, $b_2$, $c_1$ and $c_2$, i.e.~the eigenstate with $ib_1c_- = 1$ and $ib_2c_+ = 1$. Outcome-dependent phase gates on modes $c_\pm$ can be realised using phase gates and exchange gates as $S_{a_1,c-} = S_{a_1,c_1}R_{c_2,c_1}$ and $S_{a_2,c+} = S_{a_2,c_2}R_{c_2,c_1}$. The overall operation is equivalent to $[T_{a_1,a_2}][T_{c_2,c_1}] \mathcal{I}_{b_1,c_-} \mathcal{I}_{b_2,c_+}$, where we have used that $R_{c_2,a_2}R_{c_1,a_1}R_{a_1,c_-}R_{a_2,c_+} = T_{a_1,a_2}T_{c_2,c_1}$. Because of the gate $T_{c_2,c_1}$, the input magic state is consumed, i.e.~the output state of $b_1$, $b_2$, $c_1$ and $c_2$ is the eigenstate with $ib_1c_1 = 1$ and $ib_2c_2 = 1$. Therefore, the overall operation is also equivalent to $[T_{a_1,a_2}] \mathcal{I}_{b_1,c_1} \mathcal{I}_{b_2,c_2}$, i.e.~initialisation operations on $b_1$, $c_1$ and $b_2$, $c_2$ followed by a $T$ gate on $a_1$ and $a_2$.

\section{Fault-tolerant operations and magic-state preparation}
\label{sec:operations}

\begin{figure*}[tbp]
\centering
\includegraphics[width=1\linewidth]{\figpath 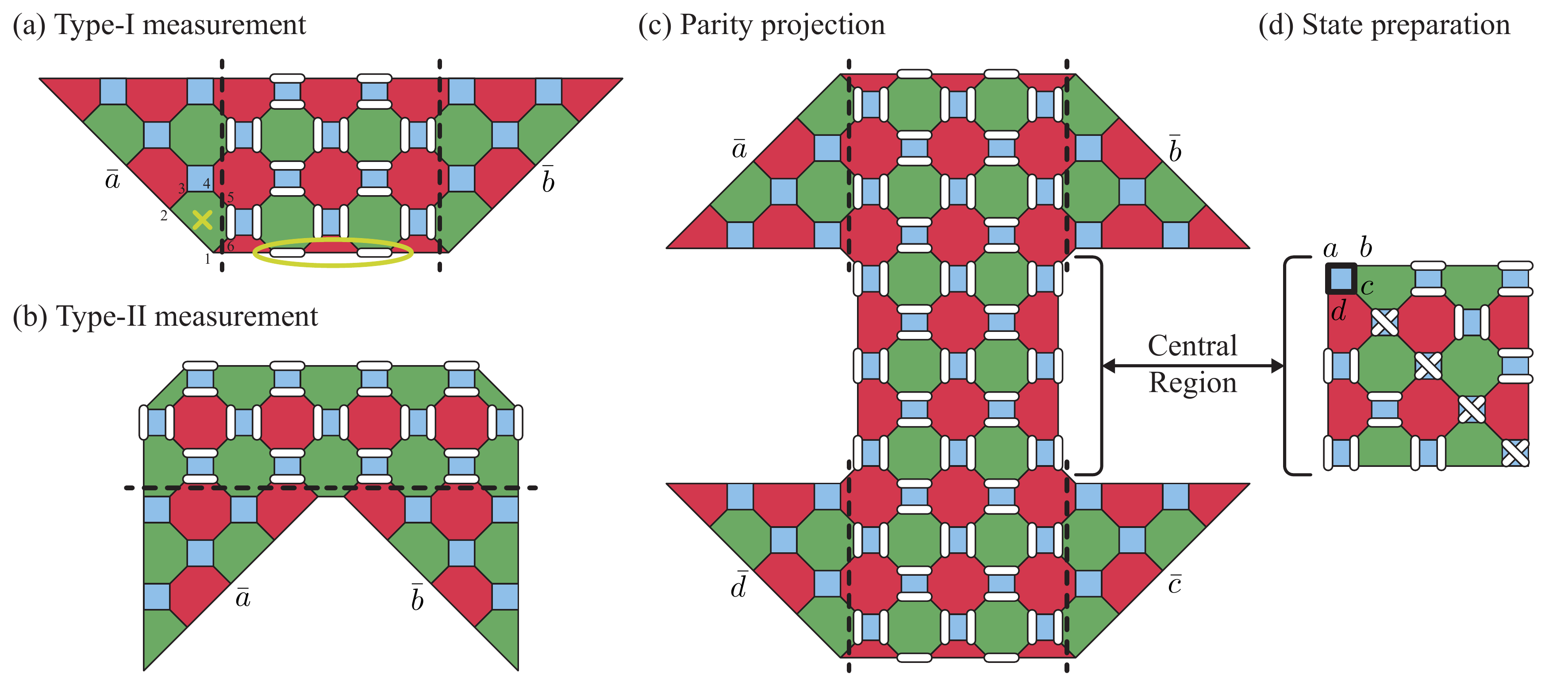}
\caption{
Operations on logical Majorana fermions. Each logical Majorana fermion mode is encoded in a triangular lattice, whose boundary is marked by the dashed line. Ancillary Majorana fermion modes are covered by white bars. Each white bar covers two vertices and denotes $ic_ic_j$, where $c_i$ and $c_j$ are two corresponding Majorana fermion operators. At the beginning of the logical operation, ancillary modes are initialised according to white bars. Then stabiliser measurements are performed on the entire lattice to read eigenvalues of stabiliser operators. Stabiliser measurements are repeated for $\sim d$ rounds. Finally, ancillary modes are measured according to white bars. In (a) and (c), the separation between dashed lines is $\sim d$. In (b), the thickness of the region above the dashed line is $\sim d$. In (d), the lattice is a square with the side length $\sim d$. Here, $d$ is the side length of logical Majorana fermions.
}
\label{fig:operations}
\end{figure*}

In this section, we discuss how to perform the phase gate, initialisation, measurement, parity projection, and preparation of magic states on logical Majorana fermions.

\subsection{Phase gate}

The phase gate $[ab]$ is equivalent to two single-mode phase operations $[a][b]$. The operation $[a]$ is not physical in a closed system, because it changes the parity of the number of fermions in the system, which is a conserved quantity in a closed system. We can realise $[a]$ by introducing ancillary Majorana fermion modes to play the role of the environment and then exchanging fermions between the system and environment. For this purpose, ancillary modes can be initialised in any state, and the overall input state is a product state $\rho = \rho_{\rm S} \otimes \rho_{E}$. Here, $\rho_{\rm S}$ is the state of Majorana fermions carrying the quantum information, and $\rho_{\rm E}$ is the state of ancillary modes. We suppose that $a$ ($b$) is an information (ancillary) Majorana fermion. After performing the phase gate $[ab]$, we get the overall output state $[ab]\rho = ([a]\rho_{S})\otimes([b]\rho_{E})$, and the output state of information Majorana fermions is $\Tr_{\rm E}([ab]\rho) = [a]\rho_{S}$. In this way, we can realise the single-mode phase operation $[a]$.

The single-mode phase operation can be performed on a logical Majorana fermion using a sequence of phase operations, i.e.~$[\bar{c}] = \prod_{i=1}^{2n-1} [c_i]$. We only need to operate one ancillary mode, because the number of modes in $\bar{c}$ is odd and each pair of single-mode phase operations is equivalent to a phase gate, i.e.~$[c_i][c_j] = [c_ic_j]$. Such a logical single-mode phase operation commutes with stabiliser operators, therefore it is a fault-tolerant operation. Using logical single-mode phase operations, we can realise the phase gate on logical Majorana fermions, i.e.~$[\bar{a}\bar{b}] = [\bar{a}][\bar{b}]$.

\subsection{Initialisation}

The initialisation can be realised using a non-destructive measurement and the single-mode phase operation. The initialisation operation reads $\mathcal{I}_{a,b} = \sum_{\nu = \pm 1} [a^\frac{1-\nu}{2}] [\pi^{(iab)}_\nu]$, where $[\pi^{(iab)}_\nu]$ corresponds to the non-destructive measurement, and $[a^\frac{1-\nu}{2}]$ is a single-mode phase operation depending on the measurement outcome.

Next, we demonstrate how to perform the non-destructive measurement on two logical Majorana fermion modes using the lattice surgery. The lattice surgery is a type of protocols for performing fault-tolerant operations on logical qubits in topological codes~\cite{Horsman2012, Landahl2014}. In this paper, we propose a set of lattice-surgery protocols to operate logical Majorana fermions. These protocols cannot be derived from lattice-surgery protocols for qubits, because each initialisation and measurement is performed on two vertices rather than one vertex (as in the qubit lattice surgery) on the color code lattice.

\subsection{Measurement}

There are two types of logical measurements. See Fig.~\ref{fig:operations}(a,b) for the protocols. The lattice of a logical Majorana fermion is a right triangle. If legs of two logical Majorana fermions face each other, we use the type-I measurement; if hypotenuses face each other, we use the type-II measurement.

In order to implement a logical measurement, we need to introduce some ancillary physical Majorana fermion modes. These ancillary modes are initialised at the beginning and measured at the end of the logical measurement. Between initialisation and measurement on ancillary modes, stabiliser operators are repeatedly measured.

The initialisation and measurement pattern are indicated by white bars in Fig.~\ref{fig:operations}(a,b). Each white bar represents a product of two Majorana fermion operators. The pattern is designed to make sure that stabiliser operators of logical Majorana fermions are not damaged in the logical measurement. For example, we consider the stabiliser operator marked by a cross in Fig.~\ref{fig:operations}(a). The stabiliser operator is $S^{(6)} = -ic_1c_2c_3c_4c_5c_6$, in which $c_1$, $c_2$, $c_3$ and $c_4$ belong to the logical Majorana fermion $\bar{a}$, and $c_5$ and $c_6$ are ancillary Majorana fermions. We note that $S^{(4)} = -c_1c_2c_3c_4$ is a stabiliser operator of $\bar{a}$. Ancillary modes are initialised in the eigenstate with $S^{(2)} = ic_5c_6 = 1$, so at the beginning of the logical measurement we have $S^{(6)} = S^{(4)}S^{(2)} = S^{(4)}$. At the end of the logical measurement, the eigenvalue of $S^{(2)}$ is measured. Then, we have $S^{(4)} = \pm S^{(6)}$, where the sign is determined by the measurement outcome $S^{(2)} = \pm 1$. We can find that, although the eigenvalue of $S^{(4)}$ may be changed by the logical measurement, we can always track its eigenvalue. Therefore, such a stabiliser operator can always be used to detect errors, which is required by the fault tolerance.

Both types of measurements are non-destructive. For the type-I measurement on logical Majorana fermions $\bar{a}$ and $\bar{b}$, we can find that $Q_{\rm R} = i\bar{a}\bar{b}Q_{\rm A}$. Here, $Q_{\rm R}$ is the product of all red stabiliser operators, which is also the product of all Majorana fermion operators; and $Q_{\rm A}$ is the product of all ancillary Majorana fermion operators, which is also the product of all white bars in Fig.~\ref{fig:operations}(a). Given values of $Q_{\rm R}$ and $Q_{\rm A}$, we can obtain the value of $i\bar{a}\bar{b}$, which is the purpose of the logical measurement. The value of $Q_{\rm R}$ can be obtained by measuring red stabiliser operators. The value of $Q_{\rm A}$ is determined by the initialisation (measurement) on ancillary modes at the beginning (end) of the logical measurement. We suppose that $Q_{\rm A} = 1$ at the beginning, then $i\bar{a}\bar{b} = Q_{\rm R}Q_{\rm A} = Q_{\rm R}$ is the input value of $i\bar{a}\bar{b}$, i.e.~the outcome of the logical measurement. Similarly, we also have the output value of $i\bar{a}\bar{b}$ at the end, which is $i\bar{a}\bar{b} = \eta_{\rm A}Q_{\rm R}$. Here, $\eta_{\rm A}$ is the product of measurement outcomes of all white bars. The output value may be different from the input value. If they are different, we can change the output value of $i\bar{a}\bar{b}$ using a logical phase operation to make sure that the measurement is non-destructive.

For the type-I measurement, the overall operation is described by the superoperator $\mathcal{M} = [\bar{a}^{\frac{1-\eta_{\rm A}}{2}}] [\pi_{\rm A}] [\pi_{\rm S}] \mathcal{I}_A$. Here, $\mathcal{I}_A$ denotes the initialisation on ancillary modes, and $\mathcal{I}_A = [\frac{1+Q_{\rm A}}{2}] \mathcal{I}_A$; $[\pi_{\rm S}]$ denotes stabiliser measurements, and $[\pi_{\rm S}] = [\frac{1+\eta_{\rm R}Q_{\rm R}}{2}] [\pi_{\rm S}]$, where $\eta_{\rm R}$ is the product of measurement outcomes of all red stabiliser operators; $[\pi_{\rm A}]$ denotes the measurement on ancillary modes, and $[\pi_{\rm A}] = [\frac{1+\eta_{\rm A}Q_{\rm A}}{2}] [\pi_{\rm A}]$; and $[\bar{a}^{\frac{1-\eta_{\rm A}}{2}}]$ is a phase operation depending on $\eta_{\rm A}$. Then, we have $\mathcal{M} = [\frac{1+\eta_{\rm R}i\bar{a}\bar{b}}{2}] \mathcal{M} [\frac{1+\eta_{\rm R}i\bar{a}\bar{b}}{2}]$. Here, we have used that $[\frac{1+\eta_{\rm A}Q_{\rm A}}{2}] [\frac{1+\eta_{\rm R}Q_{\rm R}}{2}] = [\frac{1+\eta_{\rm A}Q_{\rm A}}{2}] [\frac{1+\eta_{\rm A}\eta_{\rm R}i\bar{a}\bar{b}}{2}]$. Therefore, when the value of $Q_{\rm R}$ is $\eta_{\rm R}$, the state of logical Majorana fermions is projected to the eigenstate $i\bar{a}\bar{b} = \eta_{\rm R}$. In a similar way, we can find that the type-II measurement shown in Fig.~\ref{fig:operations}(b) is also non-destructive, and the outcome of the logical measurement (i.e.~the eigenvalue of $i\bar{a}\bar{b}$) is the product of measurement outcomes of all green stabiliser operators.

The operator $i\bar{a}\bar{b}$ is a conserved quantity in the type-I measurement, i.e.~$\mathcal{M} (i\bar{a}\bar{b} \rho) = i\bar{a}\bar{b} \mathcal{M} \rho$, where $\rho$ is the input state. First, the initialisation on ancillary modes does not affect logical Majorana fermions, i.e.~$\mathcal{I}_A (i\bar{a}\bar{b} \rho) = i\bar{a}\bar{b} \mathcal{I}_A \rho$. According to the initialisation pattern, $\mathcal{I}_A \rho = Q_{\rm b} \mathcal{I}_A \rho$ (i.e.~$Q_{\rm b} = 1$), where $Q_{\rm b}$ is the product of all ancillary Majorana fermions in the circle in Fig.~\ref{fig:operations}(a), i.e.~$Q_{\rm b}$ is the product of all white bars on the bottom raw. Then, $\mathcal{I}_A (i\bar{a}\bar{b} \rho) = i\bar{a}\bar{b} Q_{\rm b} \mathcal{I}_A \rho$. Second, stabiliser measurements commute with $i\bar{a}\bar{b} Q_{\rm b}$ (which shares even number of modes with each stabiliser operator), so $[\pi_{\rm S}] \mathcal{I}_A (i\bar{a}\bar{b} \rho) = i\bar{a}\bar{b} Q_{\rm b} [\pi_{\rm S}] \mathcal{I}_A \rho$. The product of each pair of white bars on the same blue square is a stabiliser operator, which is a conserved quantity in stabiliser measurements. Because the value of such a blue stabiliser operator is initialised as $+1$ according to the initialisation pattern, its value is still $+1$ when white bars are measured. All white bars are paired except white bars on the bottom raw. Therefore, $Q_{\rm b} = \eta_{\rm A}$ after the measurement on white bars. Third, the effect of the measurement on ancillary modes is $[\pi_{\rm A}]Q_{\rm b} = \eta_{\rm A}[\pi_{\rm A}]$, so $[\pi_{\rm A}] [\pi_{\rm S}] \mathcal{I}_A (i\bar{a}\bar{b} \rho) = \eta_{\rm A}i\bar{a}\bar{b} [\pi_{\rm A}] [\pi_{\rm S}] \mathcal{I}_A \rho$. Finally, the phase $\eta_{\rm A}$ can be cancelled by the phase operation, i.e.~$[\bar{a}^{\frac{1-\eta_{\rm A}}{2}}][\pi_{\rm A}] [\pi_{\rm S}] \mathcal{I}_A (i\bar{a}\bar{b} \rho) = i\bar{a}\bar{b} [\bar{a}^{\frac{1-\eta_{\rm A}}{2}}] [\pi_{\rm A}] [\pi_{\rm S}] \mathcal{I}_A \rho$. It is similar for the type-II measurement.

\subsection{Parity projection}

\begin{table}[h]
\begin{center}
\begin{tabular}{|c|c|c|c|c||c|c|c|c|c|}
\multicolumn{5}{c}{Stabiliser Measurement} & \multicolumn{5}{c}{Parity Projection} \\
\hline
$\eta_{8,1}$  & $\eta_{2,3}$ & $\eta_{4,5}$ & $\eta_{6,7}$ & $U$ & $\eta_{\rm t}$  & $\eta_{\rm b}$ & $\eta_{\rm l}$ & $\eta_{\rm r}$ & $\mathcal{C}$ \\
\hline
$1$  & $1$ & $1$ & $1$ & $\openone$ & $1$  & $1$ & $1$ & $1$ & $[\openone]$ \\
\hline
$-1$  & $-1$ & $1$ & $1$ & $c_1c_2$ & $-1$  & $-1$ & $1$ & $1$ & $[\bar{a}\bar{d}]$ \\
\hline
$1$  & $-1$ & $-1$ & $1$ & $c_3c_4$ & $1$  & $-1$ & $-1$ & $1$ & $[\bar{d}]$ \\
\hline
$1$  & $1$ & $-1$ & $-1$ & $c_5c_6$ & $1$  & $1$ & $-1$ & $-1$ & $[\bar{a}\bar{b}]$ \\
\hline
$-1$  & $1$ & $1$ & $-1$ & $c_7c_8$ & $-1$  & $1$ & $1$ & $-1$ & $[\bar{b}]$ \\
\hline
$-1$  & $1$ & $-1$ & $1$ & $c_1c_2c_3c_4$ & $-1$  & $1$ & $-1$ & $1$ & $[\bar{a}]$ \\
\hline
$1$  & $-1$ & $1$ & $-1$ & $c_3c_4c_5c_6$ & $1$  & $-1$ & $1$ & $-1$ & $[\bar{c}]$ \\
\hline
$-1$  & $-1$ & $-1$ & $-1$ & $c_1c_2c_5c_6$ & $-1$  & $-1$ & $-1$ & $-1$ & $[\bar{a}\bar{c}]$ \\
\hline
\end{tabular}
\end{center}
\caption{
Measurement-outcome dependent operations in the stabiliser measurement on eight Majorana fermion modes and the logical parity projection. In the stabiliser measurement, the gate $U$ is performed depending on measurement outcomes $\eta_{8,1},\eta_{2,3},\eta_{4,5},\eta_{6,7}$ [see Fig.~\ref{fig:circuits}(e)]. In the logical parity projection, the operation $\mathcal{C}$ is performed depending on measurement outcomes $\eta_{\rm t},\eta_{\rm b},\eta_{\rm l},\eta_{\rm r}$. Here, $\eta_{\rm t}$ ($\eta_{\rm b}$, $\eta_{\rm l}$ and $\eta_{\rm r}$) is the product of outcomes of white bars on the top raw (bottom raw, left column and right column) in Fig.~\ref{fig:operations}(c). We note that $\eta_{\rm A} = \eta_{\rm t}\eta_{\rm b} = \eta_{\rm l}\eta_{\rm r}$, where $\eta_{\rm A}$ is the product of outcomes of all white bars. We remark that gates $U$ and $U\prod _{i=1}^{8} c_i$ are equivalent to each other, and operations $\mathcal{C}$ and $\mathcal{C}[\bar{a}\bar{b}\bar{c}\bar{d}]$ are equivalent to each other.
}
\label{table}
\end{table}

The logical parity projection shown in Fig.~\ref{fig:operations}(c) is a non-destructive measurement of $\bar{a}\bar{b}\bar{c}\bar{d}$. We use $\mathcal{P}$ to denote the overall operation of the logical parity projection. Similar to the type-I measurement, the overall operation $\mathcal{P}$ includes the initialisation on ancillary modes, stabiliser measurements, the measurement on ancillary modes, and also logical phase operations depending on measurement outcomes. These logical phase operations are listed in Table~\ref{table}. Analysing the logical parity projection in a similar way to the type-I measurement, we can find that $\mathcal{P}$ has the following two properties, $\mathcal{P} = [\bar{\pi}]\mathcal{P}[\bar{\pi}]$ and $\mathcal{P}(i\bar{a}\bar{x}\rho) = i\bar{a}\bar{x}\mathcal{P}\rho$, where $\bar{\pi} = \frac{1+\eta_{\rm R}\bar{a}\bar{b}\bar{c}\bar{d}}{2}$, $\eta_{\rm R}$ is the product of measurement outcomes of all red stabiliser operators, and $x = b,c,d$. Therefore, the logical parity projection projects the state of four logical Majorana fermion modes to the subspace with the eigenvalue $\bar{a}\bar{b}\bar{c}\bar{d} = \eta_{\rm R}$.

We can show that $\mathcal{P}$ is equivalent to the parity projection $[\bar{\pi}]$. Similar to the state of qubits that can be expressed as a polynomial of Pauli operators, the state of fermions can be expressed using fermion operators. The input state $\rho$ can always be written as $\rho = (\alpha_{1} \otimes \pi_{\rm L} + \alpha_{a} \otimes \bar{a} \pi_{\rm L} + \ldots + \alpha_{iab} \otimes i\bar{a}\bar{b} \pi_{\rm L} \ldots) \otimes \rho_{\rm E}$, which describes three subsystems. The first subsystem is formed by ancillary modes, whose input state is $\rho_{\rm E}$. Majorana fermions for encoding $\bar{a}$, $\bar{b}$, $\bar{c}$ and $\bar{d}$ form the second subsystem, and $\pi_{\rm L}$ is the projector to the corresponding logical subspace. All other Majorana fermions form the third subsystem, described by operators $\alpha_{1},\alpha_{a},\alpha_{iab},\ldots$. The parity projection always projects the input state to a state that can be expressed as $[\bar{\pi}] \rho = (\beta_{1} \otimes \bar{\pi} \pi_{\rm L} + \beta_{iab} \otimes i\bar{a}\bar{b} \bar{\pi} \pi_{\rm L}  + \beta_{iac} \otimes i\bar{a}\bar{c} \bar{\pi} \pi_{\rm L} + \beta_{iad} \otimes i\bar{a}\bar{d} \bar{\pi} \pi_{\rm L}) \otimes \rho_{\rm E}$, where $\beta_{1},\beta_{iab},\ldots$ are functions of $\alpha_{1},\alpha_{a},\alpha_{iab},\ldots$. Next, we will show that $\mathcal{P} \rho = (\beta_{1} \otimes \bar{\pi} \pi_{\rm L}' + \beta_{iab} \otimes i\bar{a}\bar{b} \bar{\pi} \pi_{\rm L}'  + \beta_{iac} \otimes i\bar{a}\bar{c} \bar{\pi} \pi_{\rm L}' + \beta_{iad} \otimes i\bar{a}\bar{d} \bar{\pi} \pi_{\rm L}') \otimes p\pi_{\rm A}$, i.e.~the state of logical Majorana fermions is transformed as the same as in the operation $[\bar{\pi}]$. Here, $\pi_{\rm A}$ is a projector describing the final measurement on ancillary modes, $\pi_{\rm L}'$ is the projector to a new logical subspace, and $p$ is a real number. Because the projector to the logical subspace ($\pi_{\rm L}$ or $\pi_{\rm L}'$) and the state of ancillary modes do not carry any logical information, $[\bar{\pi}]$ and $\mathcal{P}$ are equivalent as operations on logical Majorana fermions.

Now, we analyse the effect of $\mathcal{P}$ on the input state. At the end of $\mathcal{P}$, all ancillary Majorana fermions are measured and projected to an eigenstate of all white bars (denoted by the projector $\pi_{\rm A}$). Similar to the type-I measurement, eigenvalues of stabiliser operators may be changed in the logical operation, i.e.~the second subsystem is projected to a new logical subspace $\pi_{\rm L}'$ at the end of $\mathcal{P}$. Therefore $\mathcal{P}(\bar{\pi} \pi_{\rm L} \otimes \rho_{\rm E}) = \bar{\rho} \pi_{\rm L}' \otimes \pi_{\rm A}$, where $\bar{\rho}$ denotes a state of logical Majorana fermions. The logical subspace of four logical modes is four-dimensional, and $\bar{\pi}$ projects logical states to a two-dimensional subspace. We can express the two-dimensional subspace as $\bar{\pi} \pi_{\rm L} = \frac{1+i\bar{a}\bar{b}}{2} \bar{\pi} \pi_{\rm L} + \frac{1-i\bar{a}\bar{b}}{2} \bar{\pi} \pi_{\rm L}$, where $\frac{1\pm i\bar{a}\bar{b}}{2} \bar{\pi} \pi_{\rm L}$ are projectors to one-dimensional subspaces. Using properties of $\mathcal{P}$, we have $\mathcal{P}(\bar{\pi} \pi_{\rm L} \otimes \rho_{\rm E}) = [\frac{1+i\bar{a}\bar{b}}{2}][\bar{\pi}] (\bar{\rho} \pi_{\rm L}' \otimes \pi_{\rm A}) + [\frac{1-i\bar{a}\bar{b}}{2}][\bar{\pi}] (\bar{\rho} \pi_{\rm L}' \otimes \pi_{\rm A})$. Because $\frac{1\pm i\bar{a}\bar{b}}{2} \bar{\pi} \pi_{\rm L}'$ are also projectors to one-dimensional subspaces, we have $[\frac{1\pm i\bar{a}\bar{b}}{2}][\bar{\pi}] (\bar{\rho} \pi_{\rm L}' \otimes \pi_{\rm A}) = p \frac{1\pm i\bar{a}\bar{b}}{2} \bar{\pi} \pi_{\rm L}' \otimes \pi_{\rm A}$. Using properties of $\mathcal{P}$ again, we can find that the probability $p$ is the same for both signs, and $p = \Tr[ \mathcal{P}(\frac{1+i\bar{a}\bar{b}}{2} \bar{\pi} \pi_{\rm L} \otimes \rho_{\rm E}) ] = \Tr[ \mathcal{P}[i\bar{a}\bar{c}](\frac{1-i\bar{a}\bar{b}}{2} \bar{\pi} \pi_{\rm L} \otimes \rho_{\rm E}) ]  = \Tr[ \mathcal{P}(\frac{1-i\bar{a}\bar{b}}{2} \bar{\pi} \pi_{\rm L} \otimes \rho_{\rm E}) ] $. Then $\mathcal{P}(\bar{\pi} \pi_{\rm L} \otimes \rho_{\rm E}) = p \bar{\pi} \pi_{\rm L}' \otimes \pi_{\rm A}$. Using properties of $\mathcal{P}$ for the third time, we have the output state $\mathcal{P} \rho = [\bar{\pi}] \mathcal{P} [\bar{\pi}] \rho = p(\beta_{1} \otimes \bar{\pi} \pi_{\rm L}' + \beta_{iab} \otimes i\bar{a}\bar{b} \bar{\pi} \pi_{\rm L}'  + \beta_{iac} \otimes i\bar{a}\bar{c} \bar{\pi} \pi_{\rm L}' + \beta_{iad} \otimes i\bar{a}\bar{d} \bar{\pi} \pi_{\rm L}') \otimes \pi_{\rm A}$.

\subsection{Magic-state preparation}

In order to prepare a magic state in logical Majorana fermions, we propose a scheme that can transfer the magic state of four physical Majorana fermion modes $a$, $b$, $c$ and $d$ to four logical Majorana fermion modes $\bar{a}$, $\bar{b}$, $\bar{c}$ and $\bar{d}$ [see Fig.~\ref{fig:operations}(c,d)]. The magic state is in a two-dimensional subspace, and without loss of generality we suppose that the subspace is $abcd = 1$. Then, the magic state can always be expressed as $\rho_{\rm in} = \frac{1+abcd}{2}\frac{1}{2}(\openone + \gamma_{iab} iab + \gamma_{iac} iac + \gamma_{iad} iad)$, where $\gamma_{iab}$, $\gamma_{iac}$ and $\gamma_{iad}$ are real numbers. Four physical modes are prepared in the state $\rho_{\rm in}$. Four logical modes are prepared in the state $\bar{\rho}_{\rm in} = \frac{1+\bar{a}\bar{b}\bar{c}\bar{d}}{2}\frac{1+i\bar{a}\bar{d}}{2}$, which can be realised by using a type-I measurement on $\bar{a}$ and $\bar{d}$ followed by a logical parity projection and phase operations if necessary. The lattice for the magic state preparation is the same as the logical parity projection [Fig.~\ref{fig:operations}(c)]. To transfer the magic state, ancillary modes in the central region of the logical parity projection are initialised according to white bars in Fig.~\ref{fig:operations}(d), and other ancillary modes are initialised as the same as in the logical parity projection. Operations after the initialisation are also the same as in the logical parity projection.

The overall operation $\mathcal{S}$ for transferring the magic state includes the initialisation on ancillary modes (excluding $a$, $b$, $c$ and $d$), stabiliser measurements, the measurement on all ancillary modes, and logical phase operations depending on measurement outcomes. The only difference between $\mathcal{S}$ and the logical parity projection $\mathcal{P}$ is the initialisation pattern. To demonstrate that $\mathcal{S}$ can transfer the magic state, we need to use the following properties, $\mathcal{S} = [\bar{\pi}]\mathcal{S} = \mathcal{S}[\frac{1+\eta_{\rm R}'\bar{a}\bar{b}ab}{2}]$, $\mathcal{S}(i\bar{a}\bar{b}\rho) = i\bar{a}\bar{b}\mathcal{S}\rho$ and $\mathcal{S}(\bar{a}\bar{d}ad\rho) = -i\eta_{iab}\bar{a}\bar{d}\mathcal{S}\rho$. Here, $\eta_{\rm R}'$ is the product of measurement outcomes of all red stabiliser operators above the central region [see Fig.~\ref{fig:operations}(c,d)], and $\eta_{iab}$ is the measurement outcome of the white bar $iab$. Similar to the logical parity projection, the output state is projected to the subspace $\bar{\pi}$, so $\mathcal{S} = [\bar{\pi}]\mathcal{S}$. We consider the product of all red stabiliser operators above the central region $Q_{\rm R}'$. Measurements on these red stabiliser operators (which are included in $\mathcal{S}$) project the state to the subspace $\frac{1+\eta_{\rm R}'Q_{\rm R}'}{2}$. Because $Q_{\rm R}'$ commutes with all white bars in the initialisation pattern, $\mathcal{S} = \mathcal{S}[\frac{1+\eta_{\rm R}'Q_{\rm R}'}{2}]$. We note that $Q_{\rm R}' = \bar{a}\bar{b}ab Q_{\rm W}'$, where $Q_{\rm W}'$ denotes the product of white bars in the initialisation pattern that covered by $Q_{\rm R}'$. Then, $Q_{\rm R}' = \bar{a}\bar{b}ab$ due to the initialisation pattern, i.e.~$\mathcal{S} = \mathcal{S}[\frac{1+\eta_{\rm R}'\bar{a}\bar{b}ab}{2}]$. Similar to the logical parity projection, $i\bar{a}\bar{b}$ is a conserved quantity in $\mathcal{S}$, so $\mathcal{S}(i\bar{a}\bar{b}\rho) = i\bar{a}\bar{b}\mathcal{S}\rho$. In the logical parity projection, $i\bar{a}\bar{d}$ is also a conserved quantity. However, because the initialisation pattern is different, $i\bar{a}\bar{d}$ is not a conserved quantity in $\mathcal{S}$, and the corresponding conserved quantity becomes $\bar{a}\bar{d}ad$, i.e.~$\mathcal{S}(\bar{a}\bar{d}ad\rho) = \bar{a}\bar{d}ad\mathcal{S}\rho$. The white bar $iab$ is measured at the end, and we have $iab = \eta_{iab}$ in the final state, so $\mathcal{S}(\bar{a}\bar{d}ad\rho) = -i\eta_{iab}\bar{a}\bar{d}\mathcal{S}\rho$.

Now, we show that the operation $\mathcal{S}$ can transfer the magic state from physical Majorana fermions to logical Majorana fermions. We suppose that the input state of all ancillary modes other than $a$, $b$, $c$ and $d$ is $\rho'$, then the output state reads $\tilde{\rho}_{\rm out} = \mathcal{S}(\bar{\rho}_{\rm in}\pi_{\rm L} \otimes \rho_{\rm in} \otimes \rho')$. Using properties of $\mathcal{S}$, we have $\tilde{\rho}_{\rm out} = \frac{1}{2}(\openone -\eta_{\rm R}' \gamma_{iab} i\bar{a}\bar{b} -\eta_{\rm R}\eta_{\rm R}'\eta_{iab} \gamma_{iac} i\bar{a}\bar{c} + \eta_{iab}\gamma_{iad} i\bar{a}\bar{d})\tilde{\rho}'$, where $\tilde{\rho}' = \frac{1}{2}\mathcal{S}(\frac{1+\bar{a}\bar{b}\bar{c}\bar{d}}{2}\pi_{\rm L}\frac{1+abcd}{2}\frac{1+\eta_{\rm R}'\bar{a}\bar{b}ab}{2} \otimes \rho')$. Similar to the parity projection, we can find that $\tilde{\rho}' = \frac{1}{2}\bar{\pi} \pi_{\rm L}' \otimes p\pi_{\rm A}$. Therefore, the magic state is transferred to logical Majorana fermions, up to phases that can be corrected using logical phase operations depending on $\eta_{\rm R}$, $\eta_{\rm R}'$ and $\eta_{iab}$.

\subsection{Errors in logical operations}

Because of the initialisation and measurement pattern, some stabiliser operators cannot be used to detect errors temporarily. We take the type-I measurement as an example [see Fig.~\ref{fig:operations}(a)], and we focus on the region between two dashed lines. At the beginning, after the initialisation on ancillary modes, the value of each green stabiliser operator is determined (which is a product of four white bars), so its outcome in the first round of stabiliser measurements can be compared with its initial value to detect errors. It is similar for blue stabiliser operators. However, for red stabiliser operators, their values cannot be determined by the initialisation pattern, and their outcomes are random in the first round of stabiliser measurements. As a result, red stabiliser operators cannot be used to detect errors at the first round of stabiliser measurements. Similarly, at the end, after the measurement on ancillary modes, we can read values of green and blue stabiliser operators from measurement outcomes of white bars but cannot read values of red stabiliser operators. As a result, we can compare final values of green and blue stabiliser operators with outcomes in the last round of stabiliser measurements to detect errors, but it does not work for red stabiliser operators. Therefore, at the beginning and the end, the examinations of red stabiliser operators are incomplete and cannot provide any information about errors. It is similar for the type-II measurement and the logical parity projection.

Errors may cause incorrect outcomes in stabiliser measurements. We consider a sequence of incorrect measurement outcomes of a stabiliser operator from the beginning to the end. These incorrect outcomes cannot be detected if there are two incomplete stabiliser examinations at the beginning and the end (i.e.~we know neither the initial value nor the final value of the stabiliser operator). Such a sequence of incorrect outcomes causes an incorrect outcome of the logical measurement or parity projection, which can be suppressed by repeating stabiliser measurements. The probability of such a logical incorrect outcome decreases with the repetition number of stabiliser measurements between two incomplete stabiliser examinations, the number of incorrect outcomes in the sequence.

The logical phase gate, initialisation, two types of measurements and parity projection are fault-tolerant operations. Logical errors in these operations can be suppressed by enlarging logical Majorana fermions, because the minimum length of nontrivial string operators is $\sim d$~\cite{Bombin2006}. A non-trivial string operator is a string operator connecting boundaries of the code. We remark that incomplete stabiliser examinations are also boundaries. 

In the magic-state preparation, the diagonal line of the square lattice [see Fig.~\ref{fig:operations}(d)] separates complete and incomplete stabiliser examinations. According to the initialisation pattern, red (green) stabiliser operators above (below) the diagonal line can be used to detect errors, but red (green) stabiliser operators blew (above) the diagonal line cannot be used to detect errors. This structure makes sure that only errors close to the top-left corner (marked by the bold black square) can cause errors on the output state of logical Majorana fermions~\cite{Li2015}, i.e.~the rate of errors on the logical magic state is $\mathcal{O}(1)$ with respect to $d$.

\section{Logical Majorana fermion array and magic-state distillation}
\label{sec:array}

\begin{figure}[tbp]
\centering
\includegraphics[width=1\linewidth]{\figpath 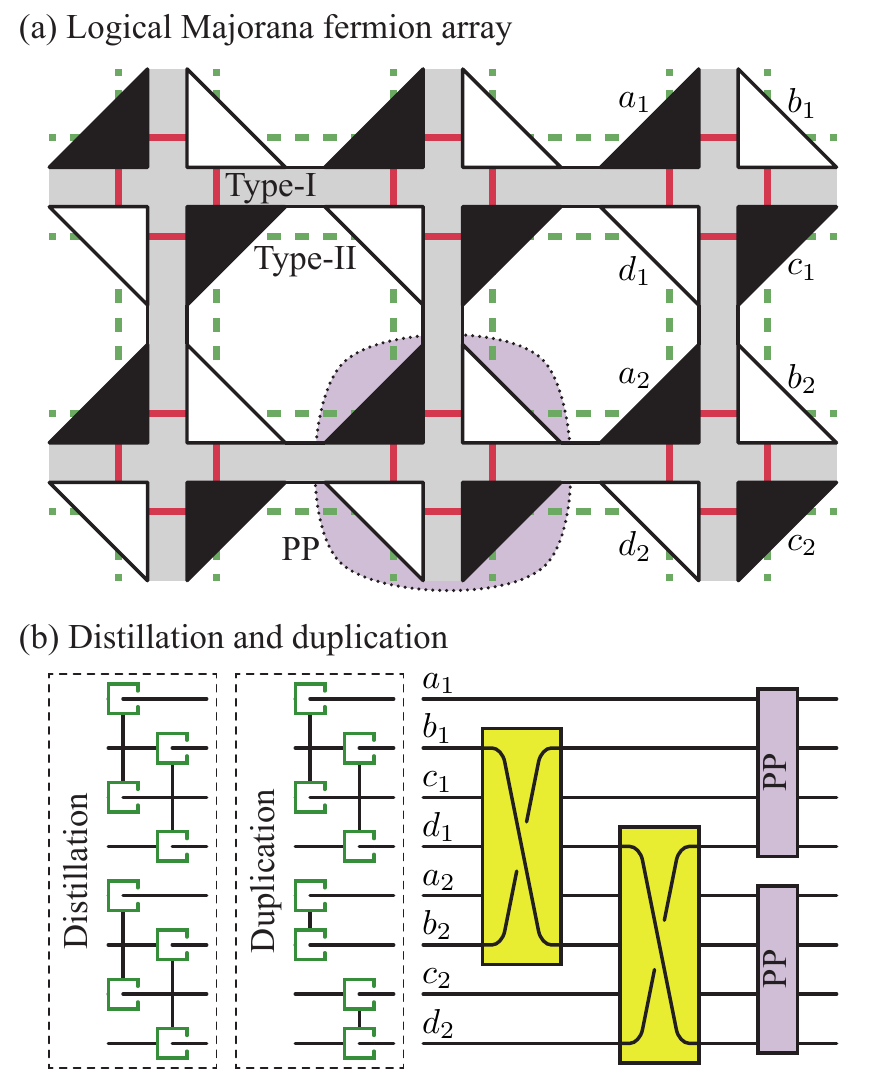}
\caption{
(a) Array of logical Majorana fermions. Each right triangle represents a logical Majorana fermion. The red solid line denotes the type-I measurement, and the green dashed line denotes the type-II measurement. The parity projection (PP) can be performed on a group of four logical Majorana fermions (e.g.~these four enveloped by the dotted line). The magic state can be prepared on such four logical Majorana fermions. The region in gray are filled by ancillary Majorana fermions. The dimension in the figure does not reflect the actual dimension on the color code lattice. (b) Circuits for the magic state distillation and duplication.
}
\label{fig:array}
\end{figure}

In addition to the set of universal operations, the universal quantum computation also requires the universal connectivity. Any parity-preserving unitary operator on the state of fermions can be achieved using the set of universal operations~\cite{Bravyi2002}, under the assumption that a $k$-mode operation can be performed on any set of $k$ Majorana fermion modes. Alternatively, we need the ability to transfer the state between any pair of Majorana fermion modes. This requirement leads to the second type of magic states.

We consider the two-dimensional array of logical Majorana fermions shown in Fig.~\ref{fig:array}(a). Connected by type-I and type-II measurements, logical Majorana fermions form a network. The state of logical Majorana fermions can be transferred among the network with the state transfer operation shown in Fig.~\ref{fig:circuits}(b), which can be realised by using logical measurements to initialise and measure logical Majorana fermions. However, the logical measurement can only be performed on a white logical Majorana fermion and a black logical Majorana fermion [see Fig.~\ref{fig:array}(a)]. In the circuit of the state transfer operation shown in Fig.~\ref{fig:circuits}(b), if $a$ is a black (white) Majorana fermion, $b$ must be a white (black) Majorana fermion, and $c$ is also a black (white) Majorana fermion. In other words, the state of a black (white) Majorana fermion can only be transferred to another black (white) Majorana fermion. Therefore, type-I and type-II measurements are not enough for the universal state transfer, and the state transfer is only allowed within a subset of logical Majorana fermions.

In order to transfer the state between two subsets of logical Majorana fermions, we have to introduce the second type of magic states. For four logical Majorana fermions $a_1$, $b_1$, $c_1$ and $d_1$ in Fig.~\ref{fig:array}(a), the magic state is the eigenstate with $ia_1c_1 = 1$ and $ib_1d_1 = 1$. Such a sate corresponds to the qubit magic state $\ket{\rm Y} = \frac{1}{\sqrt{2}}(\ket{0}+i\ket{1})$ (i.e.~$a_1b_1c_1d_1 = 1$, $\sigma^{\rm x} = ia_1d_1$ and $\sigma^{\rm z} = ic_1d_1$). Because both $a_1$ and $c_1$ ($b_1$ and $d_1$) are black (white) Majorana fermions, this magic state cannot be prepared using logical measurements. Using $a_1$ and $c_1$ in the magic state to respectively replace $b$ and $c$ initialised in $ibc = 1$ in Fig.~\ref{fig:circuits}(b), we can transfer the state of a white Majorana fermion $a$ to a black Majorana fermion $c_1$. Here $a$ is a white Majorana fermion so that the logical measurement can be performed. Similarly, using $b_1$ and $d_1$ in the magic state, we can transfer the state of a black Majorana fermion to a white Majorana fermion. Therefore, this type of magic states completes the universal state transfer required by the universal quantum computation.

Two types of magic states are required, which respectively correspond to qubit states $\ket{A}$ and $\ket{Y}$. Both of them can be prepared using the protocol shown in Fig.~\ref{fig:operations}(d). The prepared magic states are not fault-tolerant, because logical errors in these raw magic states cannot be suppressed by enlarging logical Majorana fermions. In order to obtain magic states with a high fidelity on the fault-tolerant level, we need to distil magic states.

The circuit for the distillation of Y-type magic states is shown in Fig.~\ref{fig:array}(b). Each round of the distillation needs two copies of the magic state, e.g.~prepared on $a_1,b_1,c_1,d_1$ and $a_2,b_2,c_2,d_2$ [see Fig.~\ref{fig:array}(a)], respectively. Prepared in the magic state, logical Majorana fermions $a_j,b_j,c_j,d_j$ is always in the subspace $a_jb_jc_jd_j = 1$, where $j=1,2$. We can make sure that the state is in the proper subspace using the logical parity projection, which is fault-tolerant. In this subspace, the state is either in the correct eigenstate $ia_jc_j = 1$ or the incorrect eigenstate $ia_jc_j = -1$. We remark that $ib_jd_j = ia_jc_j$ when $a_jb_jc_jd_j = 1$. We suppose that the state is in the incorrect eigenstate with the probability $p$. In the distillation circuit, we exchange $b_1$ with $b_2$ and $d_1$ with $d_2$ in order to measure $a_1b_2c_1d_2$ and $a_2b_1c_2d_1$ using logical parity projections. These exchange gates are performed on logical Majorana fermions in the same subset, so they can be realised using type-I and type-II measurements, i.e.~these exchange gates are fault-tolerant. By measuring $a_1b_2c_1d_2$ and $a_2b_1c_2d_1$, we can detect errors. If either $ia_1c_1 = -1$ or $ia_2c_2 = -1$, we have $a_1b_2c_1d_2 = -1$ and $a_2b_1c_2d_1 = -1$. Both copies of the magic state are discarded in this case. If both $ia_1c_1 = -1$ and $ia_2c_2 = -1$, we have $a_1b_2c_1d_2 = 1$ and $a_2b_1c_2d_1 = 1$, i.e.~in this case errors cannot be detected. Therefore, after one round of the distillation, the error probability is reduced from $p$ to $p^2/(1-2p+2p^2)$.

Once a copy of the high-fidelity Y-type magic state is obtained, we can duplicate the magic state as shown in Fig.~\ref{fig:array}(b). If we have only one copy of the magic state prepared on $a_1,b_1,c_1,d_1$, after the exchange gates and parity projections, we can obtain two copies of the magic state on $a_1,b_1,c_1,d_1$ and $a_2,b_2,c_2,d_2$, respectively.

Provided distilled Y-type magic states, we can implement fault-tolerant qubit Clifford gates by encoding each qubit in four logical Majorana fermions~\cite{Bravyi2006}. A-type magic states of Majorana fermions are also magic states of qubits, which can be distilled using qubit Clifford gates~\cite{Bravyi2005}.

\section{Error correction}
\label{sec:correction}

\begin{figure}[tbp]
\centering
\includegraphics[width=1\linewidth]{\figpath 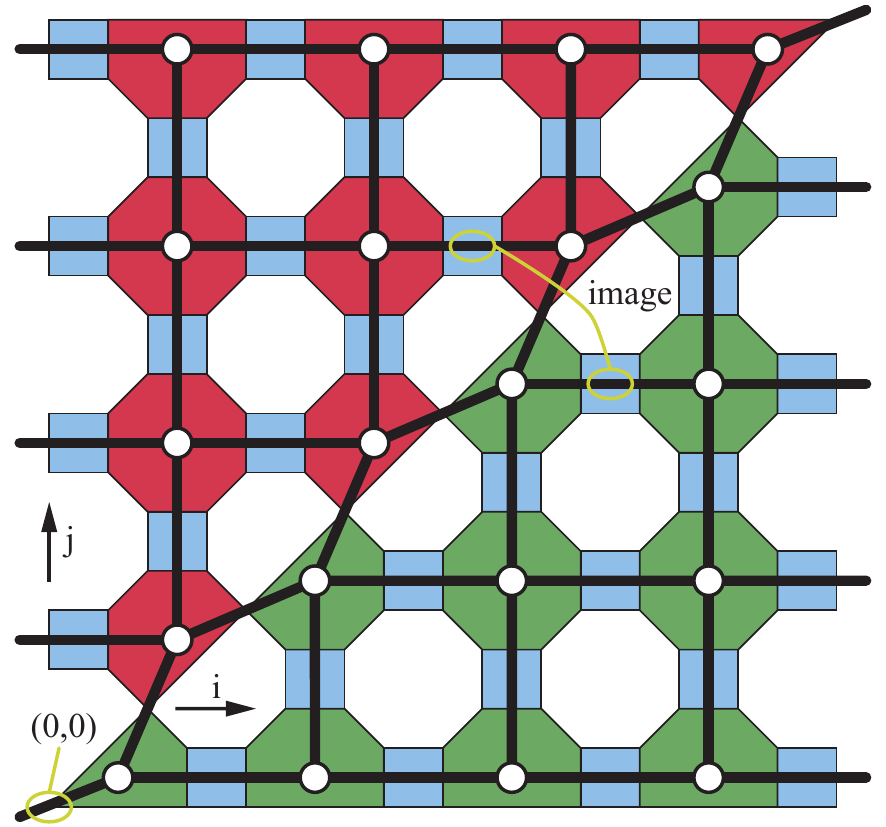}
\caption{
Lattice formed by stabiliser operators and errors. Each circle denotes a stabiliser operator, and each edge denotes an error that flips two corresponding red and green stabiliser operators but does not flip any blue stabiliser operator.
}
\label{fig:lattice}
\end{figure}

Before we discuss the decoding algorithm, we would like to firstly show that the code distance is the side length $d$ of the triangular lattice in Fig.~\ref{fig:code}. In the Majorana fermion system, the analogue of bit-flip and phase-flip errors on qubits is the error $[c]$, which is an unexpected single-mode phase operation. The error $[c]$ changes the parity of the number of fermions in the system, which could be a result of the exchange of fermions between the system and the environment. The code distance is the minimum number of single-mode errors in an error string $[\prod_{i\in V_E} c_i]$ that can change the logical state but cannot be detected by the stabiliser group, i.e.~the number of single-mode errors $\abs{V_E}$ is odd, but $\abs{V_E \bigcap V_p}$ is even for any stabiliser operator $S_p$.

Three kinds of errors cannot be detected by a blue stabiliser operator. They are $\mathcal{E}_{\rm R} = [c_1c_2]$, $\mathcal{E}_{\rm G} = [c_1c_4]$ and $\mathcal{E}_{\rm RG} = [c_2c_4]$ (see the inset in Fig.~\ref{fig:code}). We remark that the error $[c_1c_2c_3c_4]$ is trivial (because $c_1c_2c_3c_4$ is a stabiliser operator), and errors $[c_ic_j]$ and $[c_1c_2c_3c_4c_ic_j]$ (e.g.~$[c_1c_2]$ and $[c_3c_4]$) are equivalent. The error $\mathcal{E}_{\rm R}$ flips eigenvalues of two red stabiliser operators, the error $\mathcal{E}_{\rm G}$ flips eigenvalues of two green stabiliser operators, and the error $\mathcal{E}_{\rm RG}$ flips eigenvalues of all four surrounding stabiliser operators. We remark that single-mode errors on the diagonal line in Fig.~\ref{fig:lattice} are not on any blue square, therefore these single-mode errors also cannot be detected by blue stabilisers.

To find out the code distance, we reflect all green plaquettes along the diagonal line as shown in Fig.~\ref{fig:lattice}, i.e.~unfold the code~\cite{Bombin2012, Kubica2015}, and we obtain a deformed simple square lattice. In this lattice, each circle denotes either a red or blue stabiliser operator, and each edge denotes an error. Each edge connects two circles or a circle with the boundary, and the edge denotes an error that cannot flip any blue stabiliser operators but can flip corresponding red and green stabiliser operators (i.e.~connected circles). In other words, an edge connecting two red (green) plaquettes denotes a two-mode error $\mathcal{E}_{\rm R}$ ($\mathcal{E}_{\rm G}$), and an edge connecting a red plaquette with a green plaquette denotes a single-mode error $[c]$ on the vertex shared by two plaquettes. Edges connecting plaquettes with the boundary are similar.

For an error string formed by these edges, the number of edges determines the number of single-mode errors in the string. We call two edges corresponding to $\mathcal{E}_{\rm R}$ and $\mathcal{E}_{\rm G}$ on the same blue plaquette images of each other [see Fig.~\ref{fig:lattice}]. If both of them occur in the string, they only contribute two single-mode errors, i.e.~one two-mode error in the form $\mathcal{E}_{\rm RG} = \mathcal{E}_{\rm R}\mathcal{E}_{\rm G}$. We also call an edge representing a single-mode error (tilted edges in Fig.~\ref{fig:lattice}) the image of itself. Therefore, the number of single-mode errors in a string is $2n_{\rm e} - n_{\rm i} \geq  n_{\rm e}$, where $n_{\rm e}$ is the number of edges, and $n_{\rm i} \leq n_{\rm e}$ is the number of edges whose image is also in the string.

If a string visit each vertex for even times, the string cannot be detected by stabiliser operators. Therefore an error string across the lattice from the left-side boundary to the right-side boundary is nontrivial, because it cannot be detected but can change the logical state. We note that the number of single-mode errors in such a string is always odd. This is the only topology of a nontrivial string. The other topology of strings that visit each vertex for even times is closed loop. Each closed loop on the lattice corresponds to a product of stabiliser operators, so the error string is trivial and does not affect the logical state. The minimum number of edges in a nontrivial string is $\frac{d+1}{2}$. Therefore, the minimum number of single-mode errors in a nontrivial string (i.e.~the code distance) is not fewer than $\frac{d+1}{2}$.

The minimum number of single-mode errors in a nontrivial string is $d$, i.e.~the code distance is $d$. We label each horizontal edge with two coordinates $(i,j)$ and choose the edge at the lower-left corner as the origin (see Fig.~\ref{fig:lattice}). Because a nontrivial string connects the left-side and right-side boundaries, we can always select a subset of horizontal edges $\{ (0,j_0),(1,j_1),\ldots,(\frac{d-1}{2},j_{\frac{d-1}{2}}) \}$ in the string. According to the coordinate system, edges $(i,j)$ and $(j,i)$ are images of each other, so each pair of such edges occupies two values (one value) of the $j$-coordinate if $i\neq j$ ($i=j$). We use $n_{\rm i}'$ to denote the number of edges whose image is also in the selected subset. Because each value of the $i$-coordinate only occurs once in the selected subset, $n_{\rm i}' \leq j_{\rm max} - j_{\rm min} + 1$, where $j_{\rm max}$ ($j_{\rm min}$) is the maximum (minimum) value of the $j$-coordinate in the selected subset. Therefore, the number of vertical edges in the nontrivial string is $n_{\rm v} \geq j_{\rm max} - j_{\rm min} \geq n_{\rm i}' - 1$. The number of single-mode errors contributed by the selected subset is $d+1 - n_{\rm i}'$, and the number of single-mode errors contributed by vertical edges is not fewer than $n_{\rm v}$, so the total number is not fewer than $d$. Other horizontal edges that are not in the selected subset do not reduce the number of single-mode errors. If the image of an unselected horizontal edge is in the selected subset, the unselected edge does not change the number of single-mode errors in the string, otherwise it increases the number.

This method can also be used to analyse the fault-tolerance of logical operations.

The decoding algorithm is based on the lattice in Fig.~\ref{fig:lattice}. There are two steps to work out correction operations using outcomes of stabiliser measurements. In the first step, using measurement outcomes of blue stabiliser operators, we directly correct errors that flip blue stabiliser operators. If the outcome of a blue stabiliser operator is $-1$ (which should be $+1$ if the state is in the logical subspace), we perform a single-mode phase operation on one of four corresponding Majorana fermions. For example, we can always choose to perform the phase operation on the lower-left Majorana fermion ($c_3$ in Fig.~\ref{fig:code}). After the correction operation, we also need to update measurement outcomes of corresponding red and blue stabiliser operators. For $c_3$ in Fig.~\ref{fig:code}, we need to flip outcomes of the left red stabiliser operator and the lower green stabiliser operator. After the first step, all remaining errors can be mapped to edges in Fig.~\ref{fig:lattice}. In the second step, using measurement outcomes of red and green stabiliser operators, we can work out correction operations for these remaining errors using the minimum-weight perfect marching algorithm~\cite{Kolmogorov2009} as the same as the surface code~\cite{Dennis2002, Wang2011}.

False outcomes in stabiliser measurements can also be detected and corrected. If the measurement on a blue stabiliser operator reports a false outcome, an unnecessary phase operation will be performed, and outcomes of some surrounding red and green stabiliser operators will be updated incorrectly. Therefore, a false outcome of a blue stabiliser operator is equivalent to a single-mode error on one of four corresponding Majorana fermions and false outcomes of some surrounding red and green stabiliser operators. The single-mode error can be detected by later stabiliser measurements thus can be corrected. To correct false outcomes of red and green stabiliser operators, we need to use a three-dimensional cubic lattice in the decoding algorithm as the same as the surface code~\cite{Dennis2002, Wang2011}, i.e.~each layer of the cubic lattice is the square lattice shown in Fig.~\ref{fig:code}, and edges connecting layers represent false outcomes of red and green stabiliser operators.

\section{Fault-tolerance threshold}
\label{sec:threshold}

\begin{figure}[tbp]
\centering
\includegraphics[width=1\linewidth]{\figpath 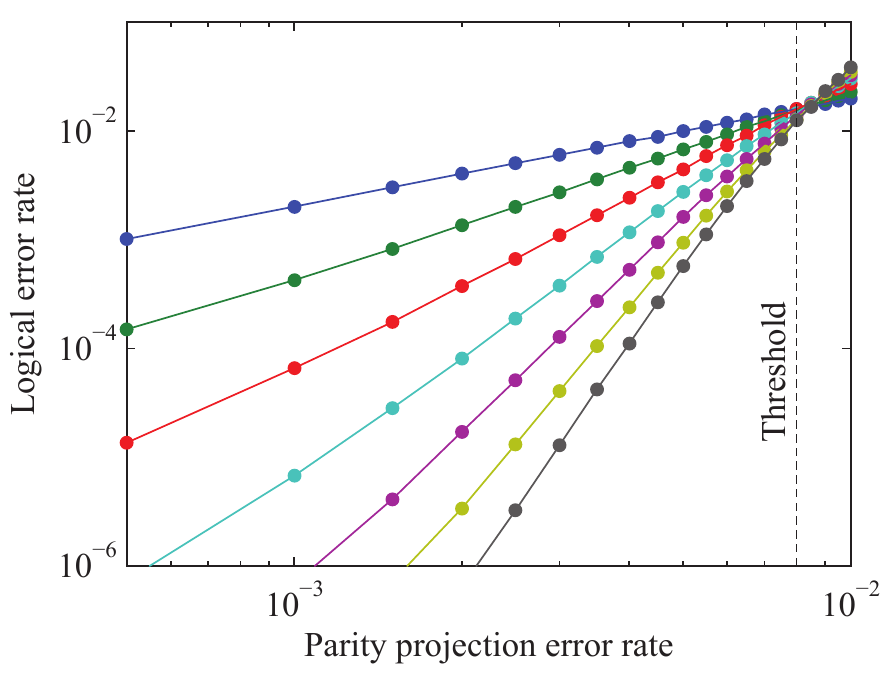}
\caption{
The rate of logical errors as a function of the rate of errors in parity projections. On the left side of the vertical dashed line, the code distance is $d = 5,9,13,17,21,25,29$ for the curves from top to bottom. Standard deviations of logical error rates are smaller than the size of circles.
}
\label{fig:plot}
\end{figure}

To study the performance of the code, we numerically simulate the error correction implemented using operations with errors. Operations used in the error correction are the initialisation, measurement and parity projection. Using these operations, we can perform stabiliser measurements: measurements on four Majorana fermion modes are performed directly using parity projections, measurements on eight Majorana fermion modes are performed using the circuit shown in Fig.~\ref{fig:circuits}(e), and measurements on six Majorana fermion modes can be performed using the same circuit by initialising two of eight modes, e.g.~$c_7$ and $c_8$, in the eigenstate $ic_7c_8 = 1$.

In our numerical simulations, we assume that two Majorana fermions $c_1$ and $c_2$ may be initialised in the incorrect eigenstate with the probability $\epsilon$, the measurement of $ic_1c_2$ may report a false outcome with the probability $\epsilon$, and for the parity projection performed on $c_1$, $c_2$, $c_3$ and $c_4$, the actual operation on the system is 
\begin{eqnarray}
\mathcal{P}_{\eta} = \mathcal{N}_{+}[\pi_{+\eta}] + \mathcal{N}_{-}[\pi_{-\eta}]
\end{eqnarray}
when the outcome of the parity projection is $c_1c_2c_3c_4 = \eta$. Here, $\pi_{\eta} = \frac{1 + \eta c_1c_2c_3c_4}{2}$ is the projector to the subspace $c_1c_2c_3c_4 = \eta$, and superoperators
\begin{eqnarray}
\mathcal{N}_{+} &=& (1-5\epsilon)[\openone] + \frac{\epsilon}{4}\sum_{i=1}^{4}[c_i] + \frac{\epsilon}{6}\sum_{i=1}^{3}\sum_{j=i+1}^{4}[c_ic_j], \\
\mathcal{N}_{-} &=& \epsilon[\openone] + \frac{\epsilon}{4}\sum_{i=1}^{4}[c_i] + \frac{\epsilon}{6}\sum_{i=1}^{3}\sum_{j=i+1}^{4}[c_ic_j].
\end{eqnarray}
The fidelity of the noisy parity projection is $1-5\epsilon$, the outcome is true but errors occur on the state with the probability $2\epsilon$, the outcome is false with the probability $\epsilon$, and a false outcome and errors on the state occur at the same time with the probability $2\epsilon$. We remark that errors $[c_i]$ and $[c_1c_2c_3c_4c_i]$ are equivalent, therefore there are only $[c_i]$ and $[c_ic_j]$ terms. When a physical Majorana fermion $c$ is waiting for operations on other Majorana fermions to be completed, memory errors in the form $[c]$ occur with the probability $\epsilon$ in the time for performing one operation on other Majorana fermions. It is reasonable to assume that the error rate of parity projections is higher than other operations, because parity projections can generate entanglements and need interactions between four Majorana fermion modes.

Numerical results are shown in Fig.~\ref{fig:plot}. The parity projection error rate is $5\epsilon$. The logical error rate in each round of stabiliser measurements is evaluated numerically using the Monte Carlo method. When the operation error rate $5\epsilon$ is below the threshold $0.8\%$, logical errors can be suppressed by enlarging the code distance.

\section{Summary}
\label{sec:summary}

In a Majorana fermion quantum computer, correcting errors using qubit error correction codes neutralises the advantage of a fermionic quantum machine in simulating other fermionic systems, and it is not necessary. We have proposed a protocol for implementing the universal fermionic quantum computation using logical Majorana fermions protected by Majorana fermion codes. Because each logical Majorana fermion is encoded in an independent set of physical Majorana fermions, logical fermionic operations are all localised, and the logical machine is a genuine fermionic quantum computer. Using the color code to protect logical Majorana fermions and the surface code decoder to correct errors, we find that the fault-tolerance threshold is as high as $0.8\%$, which can be further improved by optimising the decoding algorithm. Therefore, implementing the fault-tolerant fermionic quantum computation, which is more powerful than the fault-tolerant qubit quantum computation, requires a realistic error rate.

\begin{acknowledgments}
This work was supported by the EPSRC National Quantum Technology Hub in Networked Quantum Information Technologies. The authors would like to acknowledge the use of the University of Oxford Advanced Research Computing (ARC) facility in carrying out this work. http://dx.doi.org/10.5281/zenodo.22558.
\end{acknowledgments}

\end{document}